\documentclass[preprintnumbers,twocolumn,nofootinbib,floats,prl,superscriptaddress]{revtex4-1}

\usepackage{graphicx,amsmath,amssymb,amsfonts,hyperref}

\newcommand{\sigv}{\langle \sigma v \rangle}
\newcommand{\sutwo}{\mathrm{SU}(2)_{\rm{D}}}
\newcommand{\gD}{g_{\scriptscriptstyle \rm{D}}}
\newcommand{\lambdaP}{\lambda_{\scriptscriptstyle \rm{P}}}

\newcommand{\mhSM}{m_{h_{\rm{SM}}}}

\begin{document}

\title{\texorpdfstring{
A weighty interpretation of the Galactic Centre excess\\
\smallskip
\smallskip
\small{Phys.~ReV.~D title: Interpretation of the Galactic Center\\ excess of gamma rays with heavier dark matter particles}}{A weighty interpretation of the Galactic Centre excess}}

\preprint{IPPP/14/33 DCPT/14/64, SLAC-PUB-15945}

\author{C\'eline B\oe hm}
\affiliation{Institute for Particle Physics Phenomenology, Durham University, South Road, Durham, DH1 3LE, United Kingdom}
\affiliation{LAPTH, U. de Savoie, CNRS,  BP 110, 74941 Annecy-Le-Vieux, France}
\author{Matthew J.~Dolan}
\affiliation{Theory Group, SLAC National Accelerator Laboratory, Menlo Park, California 94025, USA\\
{\smallskip \tt  \footnotesize \href{mailto:c.m.boehm@durham.ac.uk}{c.m.boehm@durham.ac.uk}, \href{mailto:m.j.dolan@durham.ac.uk}{mdolan@slac.stanford.edu}, \href{mailto:christopher.mccabe@durham.ac.uk}{christopher.mccabe@durham.ac.uk}} \smallskip}
\author{Christopher McCabe}

\affiliation{Institute for Particle Physics Phenomenology, Durham University, South Road, Durham, DH1 3LE, United Kingdom}

\begin{abstract}

 Previous attempts at explaining the gamma-ray excess near the Galactic Centre have focussed on dark matter annihilating directly into Standard Model particles. This results in a preferred dark matter mass of 30-40~GeV (if the annihilation is into $b$ quarks) or 10~GeV (if it is into leptons). Here we show that the gamma-ray excess is also consistent with heavier dark matter particles; in models of secluded dark matter, dark matter with mass up to 76~GeV provides a good fit to the data. This occurs if the dark matter first annihilates to an on-shell particle that subsequently decays to Standard Model particles through a portal interaction. This is a generic process that works in models with annihilation, semi-annihilation or both. We explicitly demonstrate this in a model of hidden vector dark matter with an SU$(2)$ gauge group in the hidden sector.

 \end{abstract}

 \maketitle

 \section{Introduction}

Recently, an excess of gamma rays in a region of $\sim 10^{\circ}$ around the Galactic Centre has been observed in the Fermi-LAT data~\cite{Goodenough:2009gk,*Hooper:2010mq,*Boyarsky:2010dr,*Abazajian:2012pn,*Hooper:2013rwa,*Gordon:2013vta,*Abazajian:2014fta,*Daylan:2014rsa}. Although possibly consistent with astrophysical sources~\cite{Abazajian:2010zy,*Hooper:2013nhl,*Macias:2013vya,*Yuan:2014rca}, most analyses to date have focussed on interpreting the excess as a product of dark matter (DM) annihilation favouring the narrow mass range $30$-$40$~GeV for particles annihilating mainly into $b$-quarks~\cite{Boehm:2014hva,*Hardy:2014dea,*Alves:2014yha,*Berlin:2014tja,*Agrawal:2014una,*Izaguirre:2014vva,*Ipek:2014gua} or 10 GeV particles annihilating into leptons~\cite{Okada:2013bna,*Modak:2013jya,*Lacroix:2014eea,*Kong:2014haa}.
 
In this paper we emphasise that the form of the gamma-ray spectrum reflects the injection energy of the Standard Model (SM) particles from DM annihilation, rather than directly tracking the DM mass, $m_{\rm{DM}}$.  In $2\to 2$ annihilation processes that directly produce SM particles (the case that has so far been considered), the SM particles are produced with an energy $E = m_{\rm{DM}}$, leading to a direct relation between the cosmic-ray energy and the DM mass. However other modes of cosmic-ray production from DM do not feature this relation, thus allowing compatibility with DM particles over a larger mass range.

One group of examples that we highlight in this paper is secluded DM~\cite{Pospelov:2007mp} in which the DM annihilates to on-shell particle(s) $\eta$ that subsequently decay to SM particles through a portal interaction. The injection energy of cosmic rays now depends on $m_{\rm{DM}}$ and $m_{\eta}$ and the result is that DM with mass up to 76~GeV is compatible with the Fermi signal.

We demonstrate this with the secluded vector DM model proposed in~\cite{Hambye:2008bq} (see also~\cite{Hambye:2009fg,*Arina:2009uq,*Hambye:2013dgv,*Carone:2013wla,*Khoze:2014xha}). The DM in this model is three gauge bosons $Z'^a$ of the same mass that are stabilised by a custodial SO$(3)$ symmetry. The state $\eta$ is a light scalar singlet that mixes with the SM Higgs through the Higgs portal; it decays predominantly into $b$ quarks.

An attractive feature of this model is that it contains annihilation and semi-annihilation processes, which may occur when the DM is stabilised under a symmetry larger than $\mathbb{Z}_2$~\cite{D'Eramo:2010ep,*Belanger:2012vp,*D'Eramo:2012rr,*Belanger:2014bga}. This highlights that heavier DM particles may explain the Galactic Centre excess in a large class of models that have yet to be fully explored.

 \section{Hidden vector dark matter}

 The model utilised here here consists of a hidden sector with a ``dark'' SU$(2)$ gauge group (hereafter $\sutwo$) and a complex scalar doublet $\Phi$ in the fundamental of $\sutwo$. The corresponding Lagrangian is 
 \begin{equation}
 \begin{split}
 \mathcal{L}&=\mathcal{L}_{\rm{SM}}-\frac{1}{4}F_{\mu\nu}^a F^{\mu\nu a}+(D_{\mu}\Phi)^{\dagger}D^{\mu}\Phi\\
 &\qquad -\mu_{\phi}^2|\Phi|^2-\lambda_{\phi}(|\Phi|^2)^2-\lambdaP |\Phi|^2|H|^2\;, 
 \end{split}
 \end{equation}
 where $D_{\mu}=\partial_{\mu}-i \gD Z'^{a}_{\mu}t^a$, $Z'^a_{\mu}$ is the dark gauge field, $t^a$ and $F^a_{\mu\nu}$ are the $\sutwo$ generators and field strength respectively and $H$ the SM Higgs field (satisfying $\mathcal{L}_{\rm{SM}}\ni-\mu^2 |H|^2-\lambda(|H|^2)^2$).  The hidden sector communicates with the SM sector through the Higgs portal~\cite{Patt:2006fw}, with a strength determined by the $\lambdaP$ coupling.

 The electroweak and $\sutwo$ symmetries are broken by vacuum expectation values (VEVs) $v$ and $v_{\phi}$ of the fields $H$ and $\Phi$ respectively. After spontaneous symmetry breaking, we are left with two real scalars, which in the unitary gauge are
 \begin{equation}
 H=\frac{1}{\sqrt{2}}(0,v+h(x))^{\intercal};\; \Phi=\frac{1}{\sqrt{2}}(0,v_{\phi}+\phi(x))^{\intercal}\;,
 \end{equation}
 and a vector triplet $Z'^a_{\mu}$ with mass $M_{Z'}=\gD v_{\phi}/2$. The VEVs $v$ and $v_{\phi}$ satisfy the relations
 \begin{equation}
 \mu^2+\lambda v^2+\frac{\lambdaP}{2}v_{\phi}^2=0\,;\;\mu^2_{\phi}+\lambda_{\phi} v_{\phi}^2+\frac{\lambdaP}{2}v^2=0\;.
 \end{equation}

 The Higgs portal coupling $\lambdaP$ causes the $\phi$ and $h$ eigenstates to mix so that the mass eigenstates $h_{\rm{SM}}$ and $\eta$ are given by
 \begin{equation}
 \begin{pmatrix}
 h_{\rm{SM}}\\
 \eta
 \end{pmatrix}=\begin{pmatrix}
 \cos\theta&-\sin\theta\\
 \sin\theta&\cos\theta
 \end{pmatrix}
 \begin{pmatrix}
 h\\
 \phi
 \end{pmatrix}\;,
 \end{equation}
 where $\tan 2\theta=\lambdaP v v_{\phi}/(\lambda_{\phi}v_{\phi}^2-\lambda v^2)$. Two parameters among the six free parameters  $\{\mu,\mu_{\phi},\lambda,\lambda_{\phi},\lambdaP,\gD \}$ can be eliminated by requiring the observed SM Higgs boson properties ($m_{h_{\rm{SM}}}\simeq126$~GeV and $v\simeq246$~GeV) so we are left eventually with: $\{m_{\eta}, M_{Z'},\sin\theta,\gD\}$. The couplings satisfy $\gD \lesssim O(1)$ and $\gD\gtrsim\lambda>\lambda_{\phi}\sim \lambdaP$.

 \begin{figure}[t!]
 \centering
 \includegraphics[width=1.0\columnwidth]{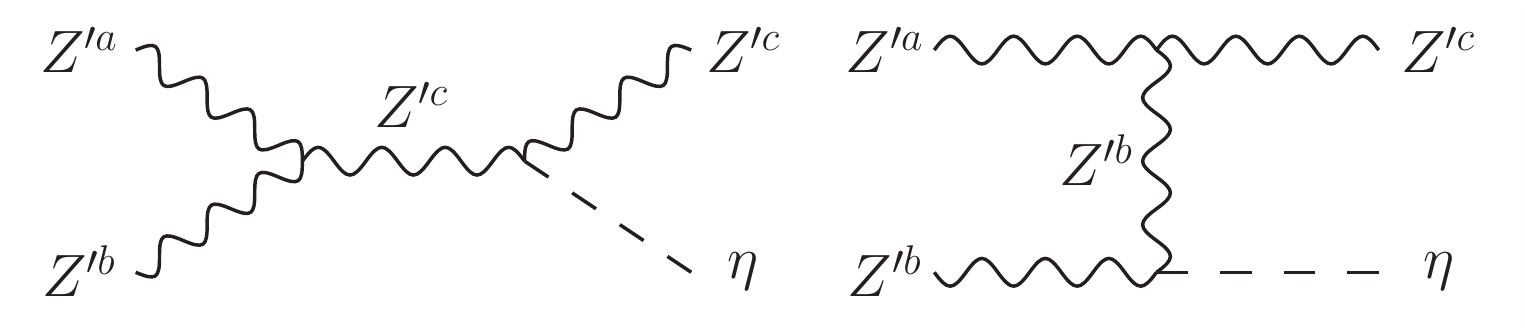}\\ \vspace{1mm}
 \includegraphics[width=1.0 \columnwidth]{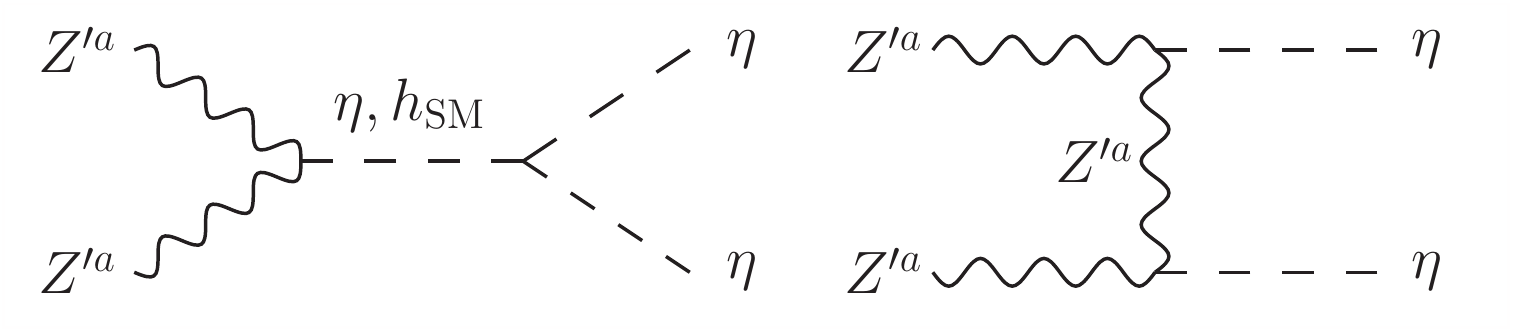}
 \caption{
 A subset of the diagrams contributing to the production of on-shell $\eta$ particles from semi-annihilation (upper diagrams) and annihilation (lower diagrams) of the DM $Z'^{a}$. The diffuse $\gamma$ spectrum arises from the decay products of $\eta\to f\bar{f}$. The direct annihilation to Standard Model particles is suppressed by $\sin^2\theta\lesssim 10^{-4}$. }
 \label{fig:Feyn}
 \end{figure}

 The three dark gauge bosons $Z'$ are stable and have the same mass due to the remnant SO$(3)$ global custodial symmetry. Hence we have three DM candidates, each of them contributing a third of the total dark mater density. All other particles are singlets under this symmetry, ensuring the $Z'$ stability.

The trilinear gauge coupling allows for semi-annihilation processes $Z'^{a} Z'^{b}\to Z'^{c} \eta$ represented in fig.~\ref{fig:Feyn}. The corresponding cross-section is given by
 \begin{equation}
 \label{eg:xsecijk}
 \langle\sigma v\rangle_{\rm{semi}}=\frac{ \gD^4 \cos^2\theta}{128 \pi M_{Z'}^4}\frac{\left( 9 M_{Z'}^4-10m_{\eta}^2 M^2_{Z'}+m^4_{\eta} \right)^{3/2}}{\left(m^2_{\eta}-3 M^2_{Z'} \right)^2}.
 \end{equation}
 We do not include semi-annihilation into $h_{\rm{SM}}$ as it is kinematically forbidden for $M_{Z'}< m_{h_{\rm{SM}}}$.

\begin{figure}[t!]
 \centering
 \includegraphics[width=0.943\columnwidth]{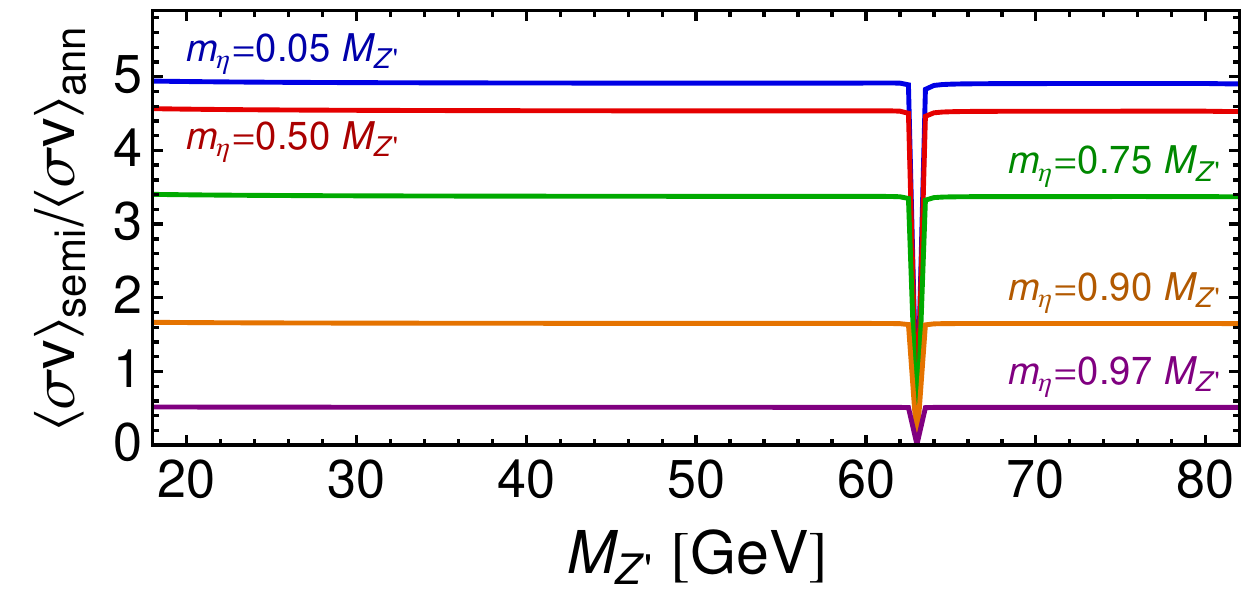}
 \caption{
 Ratio of the semi-annihilation to annihilation cross-section. We have fixed $\sin\theta=10^{-2}$ and the ratio $m_{\eta}/M_{Z'}$ to the values specified in the plot. Semi-annihilation dominates when $m_{\eta}< M_{Z'}$; annihilation dominates when $m_{\eta} \approx M_{Z'}$.
 }
 \label{fig:ratio}
 \end{figure}

The dominant annihilation channel corresponds to $Z'^{a}Z'^{a}\to \eta \eta$ and is illustrated in fig.~\ref{fig:Feyn}. The analytic expression for the corresponding cross-section ($\langle\sigma v \rangle_{\rm{ann}}$) is lengthy so we do not give it here. However, in fig.~\ref{fig:ratio}, we show the ratio $\langle\sigma v\rangle_{\rm{semi}}/\langle\sigma v \rangle_{\rm{ann}}$ for $\sin\theta=10^{-2}$ and various ratios of $m_{\eta}/M_{Z'}$.

In this plot we include all possible final states when they are kinematically accessible, including the direct $2 \to 2$ process with $Z'^{a}Z'^{a}\to \bar{f}f,W^{+}W^{-},Z^{0}Z^{0}$, even though they are suppressed by $\sin^2\theta$ (constraints discussed later require $\sin\theta\lesssim10^{-2}$). This suppression arises because $Z'^{a}$ and $\eta$ are only connected to the SM sector through $\lambdaP\propto\sin\theta$. We observe that in that regime the semi-annihilation process dominates unless $m_{\eta}\approx M_{Z'}$; annihilation eventually dominates because the phase-space suppression is faster for the semi-annihilation process. An exception is at $2M_{Z'}\approx \mhSM$ where $\langle\sigma v \rangle_{\rm{ann}}$ becomes resonantly enhanced through the diagram shown in the bottom left of fig.~\ref{fig:Feyn}. The results in fig.~\ref{fig:ratio} are unchanged for smaller values of $\sin\theta$ and furthermore, are to a very good approximation independent of $\gD$. This is because $\lambda_{\phi}\approx \gD^2 m^2_{\eta}/2 M_{Z'}^2$ for $\sin\theta \lesssim10^{-2}$, with the result that both the semi-annihilation and annihilation cross-sections scale as $\gD^4$ so the dependence of $\langle\sigma v\rangle_{\rm{semi}}/\langle\sigma v \rangle_{\rm{ann}}$ on $\gD$ drops out.

 \section{The diffuse excess}

 When annihilation to on-shell $\eta$ particles is kinematically allowed, the dominant contribution to the diffuse $\gamma$ spectrum is from the (rapid) decay of $\eta$ to SM fermions $f \bar{f}$. The prompt differential gamma-ray flux from either semi-annihilation or annihilation processes  with an on-shell $\eta$ in the final state 
 that  decays into $f \bar{f}$ is
 \begin{equation}
 \frac{d^2 \Phi}{d \Omega d E_{\gamma}}= \frac{N_{Z'} \sigv J(\tilde{\theta})}{8 \pi M_{Z'}^2}\sum_f \mathrm{BR}^{\eta \to f \bar{f}} \left(\frac{d N}{d E_{\gamma}}\right)^{Z'\!\!,f\bar{f}}.
 \end{equation}
 Here $N_{Z'}$ is a model specific combinatoric factor; $\sigv$ is the (semi)-annihilation cross-section; the $J$-factor is $J(\tilde{\theta})=\int d\lambda \rho^2(\lambda,\tilde{\theta})$, where $\lambda$ is the line of sight distance, $\tilde{\theta}$ the angle between the line of sight and the Earth-Galactic Centre axis and $\rho$ the halo profile; $\mathrm{BR}^{\eta \to f \bar{f}}$ is the branching ratio for the decay of $\eta \to f \bar{f}$ and $\left(d N/d E_{\gamma}\right)^{Z'\!\!,f\bar{f}}$ is the photon multiplicity per annihilation in the Galactic rest frame.

In the hidden vector model $N_{Z'}=\{1/3,2/3\}$ when $\sigv=\{\langle \sigma v\rangle_{\rm{semi}},\langle \sigma v\rangle_{\rm{ann}}\}$. For $\rho$ we follow~\cite{Daylan:2014rsa} by taking a generalised NFW profile~\cite{Navarro:1995iw} with $\gamma=1.26$, $r_s=20$~kpc and a normalisation giving a local density 0.3~GeV/cm$^3$ at 8.5~kpc from the Galactic Centre.  This profile is slightly cuspier than the standard choice but consistent with the results of numerical simulations~\cite{Diemand:2008in}. The branching ratios of $\eta$ decays are the same as the branching ratios of a SM-like Higgs with mass $m_{\eta}$; for the mass range we consider the dominant decay is to $b\bar{b}$. Owing to this, in this paper we do not take into account the diffuse photon emission from primary and secondary electrons as their effect is small for $b\bar{b}$ final states~\cite{Lacroix:2014eea}.

 \begin{figure}[t!]
 \centering
 \includegraphics[width=0.95\columnwidth]{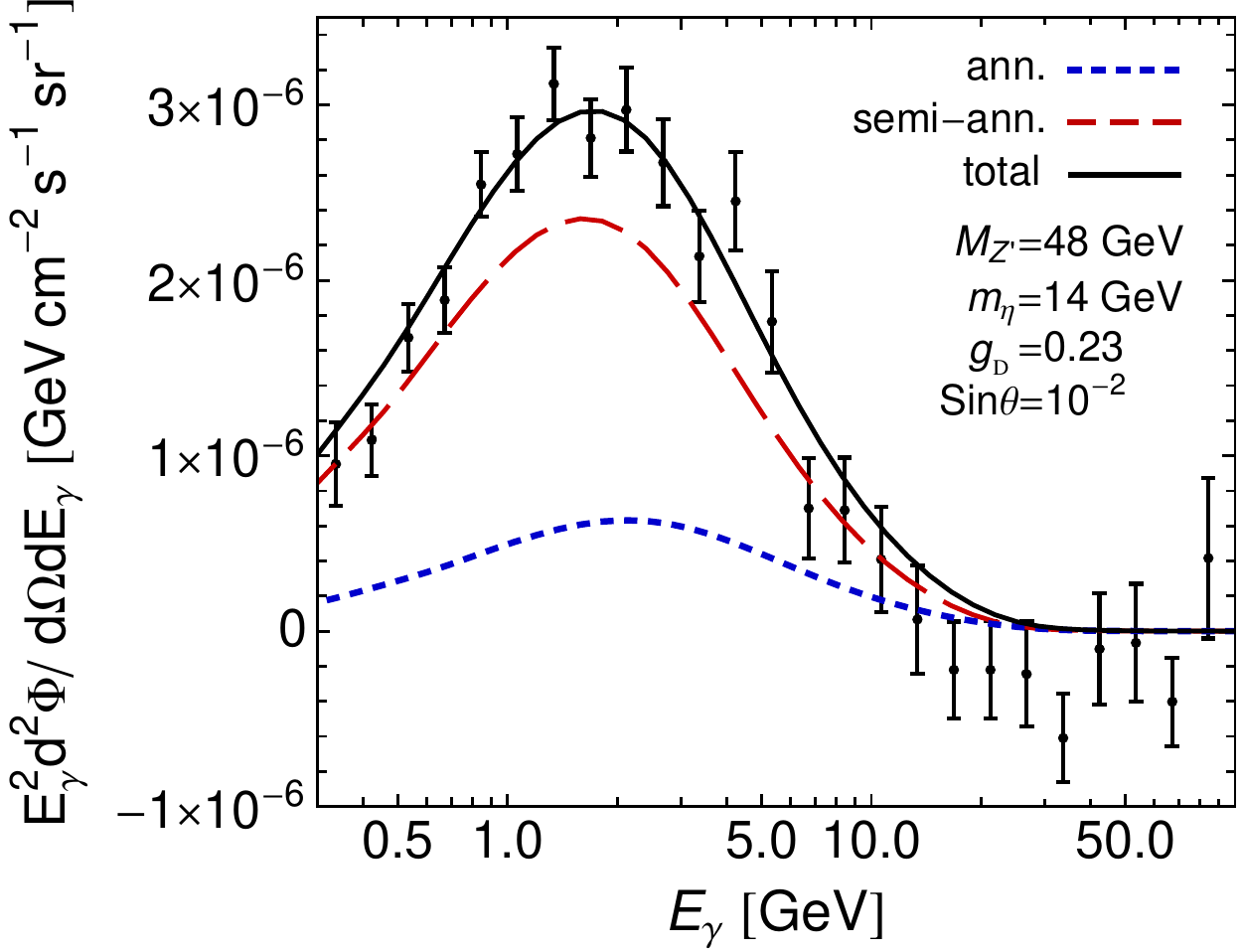}
 \caption{
The spectrum from annihilation (blue dotted), semi-annihilation (red-dashed) and their sum (black-solid) for the stated parameters. The data points show the Galactic Centre excess spectrum from the inner galaxy assuming a generalised NFW profile with $\gamma=1.26$ (from~\cite{Daylan:2014rsa}).}
 \label{fig:spectrum}
 \end{figure}

The $\eta$ particles are in general not produced at rest so we must relate the photon multiplicity in the rest frame of $\eta$, $\left(dN/dE_{\gamma}\right)^{\eta,f\bar{f}}$, to the photon multiplicity in the Galactic rest frame, $\left(d N/d E_{\gamma}\right)^{Z'\!\!,f\bar{f}}$. They are related by~\cite{Bergstrom:2008ag}
 \begin{equation}
 \left(\frac{d N}{d E_{\gamma}}\right)^{Z'\!\!,f\bar{f}}=\frac{N_{\eta}}{2 \beta \gamma}\int_{E_{\gamma}/\gamma (1+ \beta)}^{E_{\gamma}/\gamma (1- \beta)}\frac{d E'_{\gamma}}{E'_{\gamma}}\left(\frac{dN}{dE'_{\gamma}}\right)^{\eta, f\bar{f}},
 \end{equation}
 where $N_{\eta}=\{1,2\}$ for $\{$semi-annihilation, annihilation$\}$ and the boost factors $\gamma=(1-\beta^2)^{-1/2}$ are
 \begin{equation}
 \gamma_{\rm{ann}}=\frac{M_{Z'}}{m_{\eta}},\;\;\gamma_{\rm{semi}}=\frac{3 M_{Z'}^2+m_{\eta}^2}{4 M_{Z'}m_{\eta}}
 \end{equation}
 respectively. We use the values of $\left(dN/dE_{\gamma}\right)^{\eta, f\bar{f}}$ tabulated in~\cite{Cirelli:2010xx}, which were generated with {\texttt Pythia~8.135}~\cite{Sjostrand:2007gs}.

 \begin{figure}[t!]
 \centering
 \includegraphics[width=0.95\columnwidth]{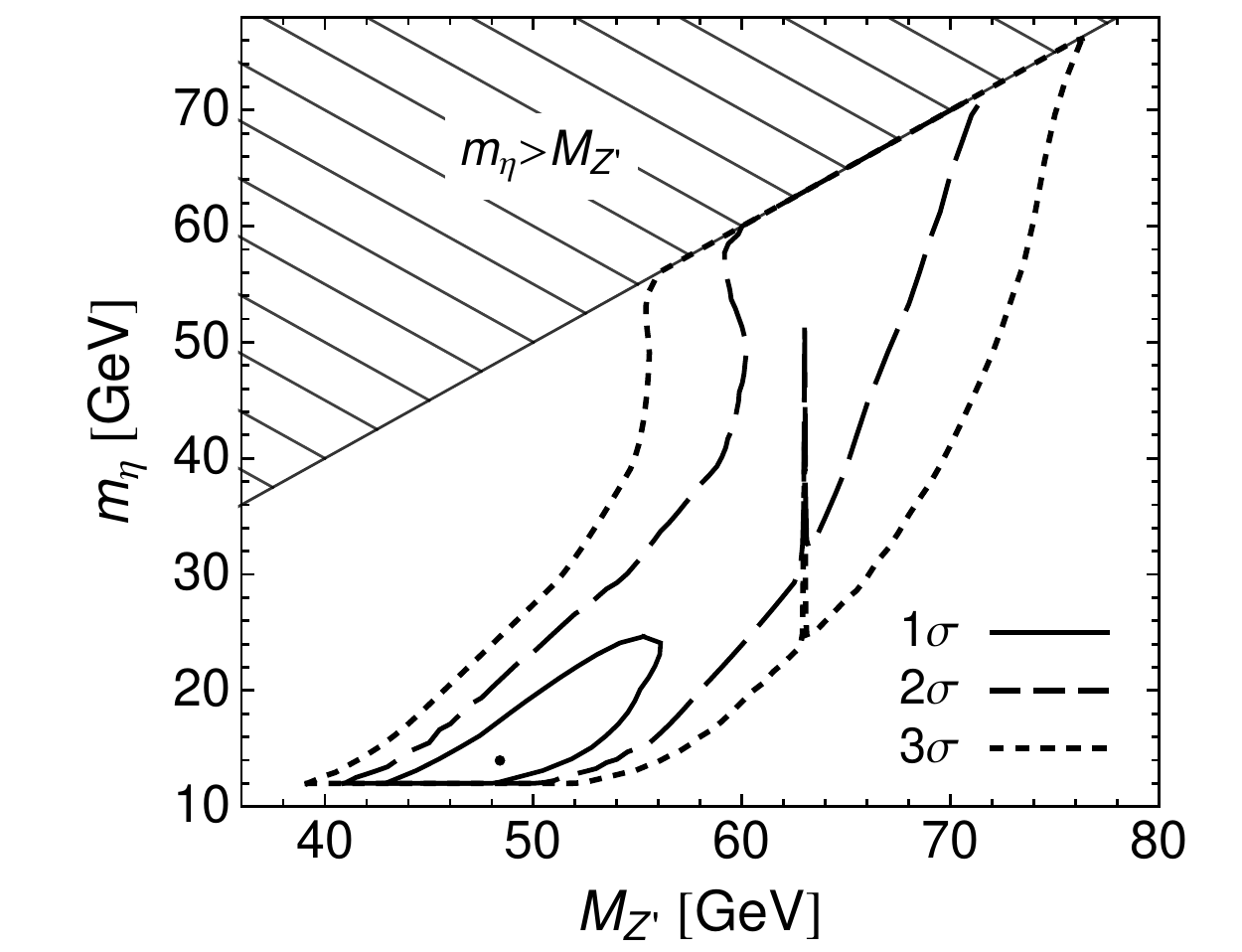}
 \caption{The 1, 2 and $3\sigma$ regions from our fit to the Galactic Centre excess. We marginalise over $\gD$ and fix $\sin\theta=10^{-2}$. The dot is the best-fit point whose spectrum is shown in fig.~\ref{fig:spectrum}. On-shell production of $\eta$ is forbidden in the hatched region. The spike is due to a resonance with $h_{\rm{SM}}$.}
 \label{fig:semiann}
 \end{figure}

To fit the Galactic Centre excess we use inner galaxy data from~\cite{Daylan:2014rsa}. They fit the Fermi data using a combination of point sources and four templates, corresponding to diffuse photon emission, an isotropic template, a template coincident with the Fermi bubbles and a DM template. The black data points and error bars in fig.~\ref{fig:spectrum} show the result from the DM template for the best-fit value $\gamma=1.26$. The spectrum has been normalised to the value of the photon flux at $\tilde{\theta}=5^{\circ}$ from the Galactic Centre.

For our results we fixed $\sin\theta=10^{-2}$ while scanning over $M_{Z'}$, $m_{\eta}$ and marginalising over $\gD$ (the results are the same for smaller values of $\sin\theta$). The red-dashed and blue-dotted lines in fig.~\ref{fig:spectrum} show the spectrum from the semi-annihilation and annihilation processes for the parameters listed in the figure, which give the best fit to the data. The cross-sections for these parameters are $\langle \sigma v \rangle_{\rm{semi}}=9.1\times10^{-26}~\text{cm}^3\,\text{s}^{-1}$ and $\langle \sigma v \rangle_{\rm{semi}}/\langle \sigma v \rangle_{\rm{ann}}=4.8$. We observe that the annihilation spectrum is slightly broader than the semi-annihilation spectrum because the $\eta$ is boosted more in the former case. The black solid line shows the total spectrum from both contributions and has $\chi^2_{\rm{min}}=29.7$, comparable to the $\chi^2$ of 28.6 found in~\cite{Daylan:2014rsa} for annihilations proportional to the square of the mass of the final state.

 Figure~\ref{fig:semiann} shows the 1, 2 and 3$\sigma$ contours in the $M_{Z'}$-$m_{\eta}$ plane after marginalising over $\gD$. The dot indicates the best-fit point and the spike at $M_{Z'}\approx63$~GeV is because of the resonant enhancement of $\langle \sigma v \rangle_{\rm{ann}}$ with $h_{\rm{SM}}$ (through the lower left diagram of fig.~\ref{fig:Feyn}). There are two hard edges to the contours: the first indicates the region where $m_{\eta}>M_{Z'}$ so that (semi)-annihilation to on-shell $\eta$ particles is forbidden. The second at $m_{\eta}=12$~GeV is where our calculation of $\left(dN/dE_{\gamma}\right)^{\eta, f\bar{f}}$ becomes unreliable. There is no physical reason for this edge; a change will occur for $m_{\eta}<2 m_b$ when decays to $b \bar{b}$ are kinematically forbidden.
 
 Another feature in fig.~\ref{fig:semiann} is narrowing of the $2\sigma$ contour when $m_{\eta}\approx50$~GeV. This narrowing is the meeting of two separate regions: In the first at $M_{Z'}\approx m_{\eta}$, $\eta$ is produced almost at rest so that the SM fermion $f$ energy is approximately $M_{Z'}/2$. We then expect that the $2\sigma$ range of $M_{Z'}$ is double the $2\sigma$ mass range found in~\cite{Daylan:2014rsa} for the standard annihilation process; this is indeed the case. In the second at $m_{\eta}\ll M_{Z'}$, $\eta$ is boosted so smaller values of $m_{\eta}$ are able to produce $f$ with energy $\sim35$~GeV in the Galactic rest frame.

Finally, it is interesting to compare the cross-section required to explain the Galactic Centre excess with that for the observed relic abundance from thermal freeze-out. Upon numerically solving the Boltzmann equation for the total abundance, we find that the observed abundance is obtained when $\tfrac{1}{3}\langle\sigma v\rangle_{\rm{ann}}+\tfrac{2}{3}\langle\sigma v\rangle_{\rm{semi}}\approx 2.5\times 10^{-26}~\mathrm{cm}^3~\!\mathrm{s}^{-1}$. We find that this combination of cross-sections in the $1\sigma \,(3\sigma)$ region is a factor $2.3-3.0\,(1.4 - 4.0)$ higher. This difference may be ameliorated when uncertainties in the astrophysical parameters are taken into account.

 \section{Other Constraints}

At direct detection experiments, $Z'^{a}$ can elastically scatter with a nucleon $N$ via exchange of $\eta$ or $h_{\rm{SM}}$; the resulting spin-independent scattering cross-section is
 \begin{equation}
 \sigma_{N}^{\rm{SI}}\approx5\times 10^{-46}~\mathrm{cm}^2 \left(\frac{\gD}{0.25}\right)^{2} \left(\frac{\sin\theta}{10^{-2}}\right)^{2} \left( \frac{25~\mathrm{GeV}}{m_{\eta}}\right)^{4}.
 \end{equation}
 The current bound from LUX $\sigma_{N}^{\rm{SI}}<8\times10^{-46}~\mathrm{cm}^2$~\cite{Akerib:2013tjd} imposes $\sin\theta\lesssim 10^{-2}$. 
 
 Contributions to the Higgs boson invisible width provide further constraints. The contribution from $\eta$ to the Higgs width is
 \begin{equation}
 \Gamma^{h_{\rm{SM}}}_{\eta\eta}\approx 0.1~\mathrm{MeV}\left(\frac{\gD}{0.5}\right)^2 \left(\frac{\sin\theta}{0.1} \right)^2  \left( \frac{50~\mathrm{GeV}}{M_{Z'}}\right)^2\;,
 \end{equation}
 where the approximation holds for small $\sin\theta$. This contribution is small compared to the total width $\sim4.2$~MeV and is unlikely to be observable in future experiments. We also found that the contribution from $h_{\rm{SM}} \to Z'^{a} Z'^{a}$ is below current limits; the best-fit parameters contribute 0.01~MeV to the width for instance. Furthermore, we have checked our scenario against direct searches for Higgs bosons at LEP~\cite{Barate:2003sz} and precision electroweak constraints~\cite{Peskin:1991sw,*Beringer:1900zz}, which constrain $\sin\theta \lesssim 10^{-1}$. While some of these constraints will tighten after the 13/14~TeV run of the LHC, we find that this scenario is unconstrained for $\sin\theta\lesssim10^{-2}$.

 Finally, there are constraints from Fermi gamma-line searches~\cite{Ackermann:2013uma}. This is relevant because $\eta$, like $h_{\rm{SM}}$, has loop-induced decays to two photons. As the $\eta$ is boosted, this decay looks like a box feature rather than a line~\cite{Ibarra:2012dw,*Ibarra:2013eda}. Owing to the small branching ratio to two photons ($\sim 10^{-4}$ for $m_{\eta}\approx30$~GeV), the flux is below current Fermi limits. For instance, the flux $\Phi_{\gamma \gamma}\approx4\times10^{-13}$~cm$^{-2}$~s$^{-1}$ for the best-fit parameters is over an order of magnitude below the limit $\Phi_{\gamma \gamma}\lesssim 5\times 10^{-11}$~cm$^{-2}$~s$^{-1}$ for the region optimised for a contracted NFW halo in~\cite{Ackermann:2013uma}.

 \section{Summary}
 
The spectrum of the Galactic Centre excess constrains the injection energy of the SM particles and not directly the mass of the DM responsible for their production. In secluded DM models in which the dominant annihilation channel is to on-shell particle(s) $\eta$ that subsequently decay to SM particles, the cosmic-ray injection energy depends on $m_{\rm{DM}}$ and $m_{\eta}$. We demonstrated that in these models, DM with mass 39-76~GeV provides a good fit to the Galactic Centre excess; this mass range is four times larger than that found previously for models in which DM annihilates directly to SM particles. The Higgs portal coupling that allows $\eta$ to decay also naturally explains why the dominant decay is into $b$ quarks, as preferred by the data. By considering a model of hidden vector DM, we demonstrated that this mechanism works for both annihilation (which dominates when $m_{\eta} \approx M_{Z'}$) and semi-annihilation (dominating when $m_{\eta} < M_{Z'}$). This paper opens up a large number of model building possibilities to explain the Galactic Centre excess beyond those that have previously been considered.


\section*{Acknowledgements}

CB thanks Joe Silk for discussions. MJD thanks Tim Cohen and Tracy Slatyer for discussions. CM thanks Dan Hooper, Valya Khoze, Matthew McCullough, Gunnar Ro and the participants of the `Bright ideas on dark matters' workshop for discussions, and CP3-Origins for hospitality while part of this work was completed.\\\\
{\bf Note added:} Ref.~\cite{Ko:2014gha}, which was listed on the arXiv on the same day as v.1 of this paper, also showed that heavier DM particles in an Abelian vector secluded model are consistent with the Galactic Centre excess.

\bibliography{ref}
\bibliographystyle{apsrev4-1}

\end{document}